# Benchmarking locally hosted language models for journal editorial work on a compact desktop workstation

**Haruka Ozaki**[1,2,3]

[1] RIKEN Center for Biosystems Dynamics Research, Kobe, Hyogo 650-0047, Japan

[2] Institute of Medicine, University of Tsukuba, Tsukuba, Ibaraki 305-8575, Japan

[3] Laboratory Automation Suppliers' Association, Kobe, Hyogo 650-0047, Japan

*Corresponding author: Haruka Ozaki, RIKEN Center for Biosystems Dynamics Research, Kobe, Hyogo 650-0047, Japan. ORCID iD: https://orcid.org/0000-0002-1606-2762. e-mail: ai-biology@ml.riken.jp*

*Running title: Locally hosted models for journal editorial work*

## Abstract

Journals are beginning to consider language models for manuscript handling, but submitted manuscripts are unpublished, and where policy forbids sending them to an external service the model must run on hardware the journal controls. The capability of locally hosted models on editorial work has not been measured. Here we constructed a benchmark of eight editorial tasks from a journal's Instructions for Authors, from manuscripts carrying defects we seeded and verified independently, and from published reviews of a preprint, and evaluated twenty open-weight models spanning a twenty-five-fold range of weight size on a compact desktop workstation of the kind a laboratory or small editorial office can adopt. The strongest model detected 36 of 40 seeded guideline violations and occupied 81 GB; a 17 GB model detected 33. Across the best configurations tested, we observed no consistent monotonic association between weight size and score: rank correlations were negligible on every task (Spearman |rho| <= 0.19), and within one model family the larger member scored below its smaller sibling. A deterministic checker of regular expressions and arithmetic, using no model, detected 31 of the same violations in a fraction of a second, and the union of its detections with those of the strongest model covered all 40. On the single peer-review case, the best model recovered 6 of 12 points from three published reviews. Prompt structure substantially altered scores within individual models. This level of performance is therefore within reach of a workstation of this class, once the deterministic checks are written.

## ◄ Significance ►

Manuscripts under review are confidential; where policy forbids sending them to an external service, the model must run on hardware the journal controls. How capable such models are has not been measured. Across eight editorial tasks on a compact desktop workstation we found no detectable size-quality association: a 17 GB model detected 33 of 40 seeded guideline violations where the largest detected 36. Three quarters of those violations were recovered by a short program using no model at all. Editorial support of this kind is within reach of a workstation of this class.

## Introduction

The volume of scholarly publication has grown exponentially for several decades, well before the advent of generative AI [1]. Recent analyses suggest that this growth accelerated following the public release of generative-AI services [2]. Similar effects have been observed at the level of individual researchers, whose output and reach increase measurably when such tools are used [3]. This trend is not merely incidental: integrating AI into the research process has been argued to offer substantial gains in scientific productivity, motivating corresponding investment in research infrastructure and policy [4].

Upstream stages of research are also becoming increasingly automated. End-to-end systems for conducting research [5], multi-agent systems that plan and execute discovery [6,7], autonomous agents coordinating through shared artifacts [8,9], tool-augmented models for domain-specific work [10], and systems that generate hypotheses from literature and data [11] have appeared in rapid succession. As these systems mature, their outputs increasingly take the form of manuscripts that enter conventional editorial workflows.

Editorial responses to generative AI have so far focused primarily on exclusion, with good reason. Generative models introduce specific and detectable defects into scholarly writing. Their bibliographic citations may be fabricated or erroneous [12]; in systematic-review settings, a substantial fraction of generated references have been found not to exist, with the majority being invalid for some models [13]. More subtly, models may cite real papers for claims those papers do not support [14]. Such errors have already entered the published record: approximately 300 papers were found to contain at least one hallucinated citation, predominantly in publications from 2025, including more than one hundred accepted conference papers [15]. Automated screening for these defects is therefore a defensible application of AI, and their detection is becoming increasingly tractable [16].

Screening, however, addresses defective submissions; it does not address the growing volume of adequate submissions implied by the trends above. The quality effects of AI assistance remain contested. Language models used as labour-augmenting technologies have been reported to increase researchers' output while reducing average quality [17], whereas other work finds simultaneous gains in productivity and quality [18]. A third line of evidence suggests that AI tools increase individual impact while narrowing the range of questions pursued by a research community [3], although deliberate sampling of under-explored directions can mitigate this narrowing [19]. These findings disagree on the direction of quality effects but agree that output increases. The editorial bottleneck is therefore likely to shift from identifying inadequate submissions to processing an increasing number of adequate ones reliably and efficiently, given that the pool of available reviewers grows much more slowly than submission volume [20,21].

Editorial support is a natural target for automation under this constraint. Checking a manuscript against submission guidelines is rule-governed and repetitive. Supporting peer review by surfacing issues that a reviewer might otherwise overlook is less rule- governed, but still bounded in scope. Such automation is not purely hypothetical: the Journal of Open Source Software already conducts substantial parts of its editorial workflow through bots acting on repository issues [22].

The principal constraint on deploying more capable systems is confidentiality. Submitted manuscripts are unpublished and may contain proprietary results or intellectual property, and publisher policies restrict their transmission to externally operated AI services. Major publishers, including Elsevier and Springer Nature, caution reviewers against uploading submitted manuscripts to AI tools because of concerns over confidentiality and proprietary rights [23,24]. It has additionally been argued that confidentiality guarantees for external services cannot be independently verified and that uploaded material may be retained for system improvement, making zero risk unattainable [25]. Although these policies are written for peer reviewers, the same confidentiality considerations would reasonably be expected to extend to editorial offices handling the same documents. These constraints therefore motivate the use of AI tools that run on hardware under the editorial office's own control.

Locally hosted language models, however, do not generally represent the performance frontier. Open-weight models that fit on consumer hardware trail proprietary frontier models by approximately six to twelve months on standard benchmarks [26], so an editorial office adopting local inference may accept some performance lag in exchange for confidentiality. Whether the resulting capability is sufficient for editorial work remains unclear. Previous work on publication-stage tasks has developed datasets and systems for scientific-claim verification, figure interpretation, and novelty assessment [27,28,29]. These studies, however, typically compare models with one another rather than asking what level of editorial capability can be obtained on locally controlled hardware. Existing reports of local inference likewise tend to characterize throughput rather than task quality in editorial settings.

This study therefore formulates editorial work as a benchmark for locally hosted language models. We ask three questions. First, how much editorial work can an open- weight model perform, and how large must the model be to do so? Second, does editorial capability scale predictably with model size, as hardware-provisioning decisions implicitly assume? Third, how strongly is performance affected by factors other than model size, including task formulation, tool access, and the hardware on which inference is performed?

To answer these questions, we constructed eight tasks from a journal's Instructions for Authors, manuscripts containing seeded and independently verified defects, and the published reviews of a preprint (Figures 1 and 2). We evaluated twenty models ranging from 3.3 to 81 GB on a single compact desktop workstation. As a deterministic baseline, the same items were also evaluated using a short program that performs the relevant checks without a language model.

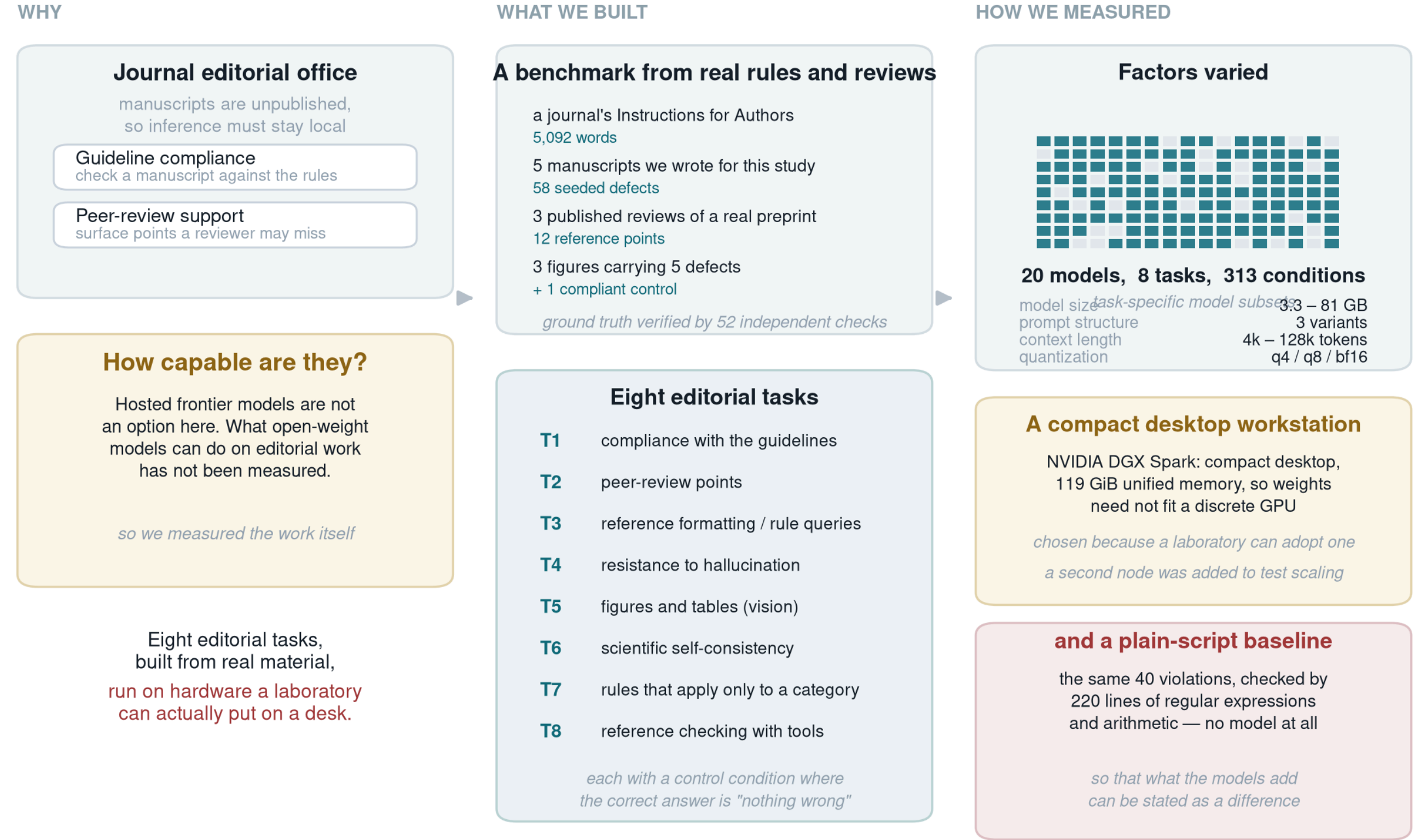

**Figure 1** The question and the design of the study. Manuscript confidentiality requires that inference run on hardware the editorial office controls, which makes the capability of locally hosted language models the relevant quantity. The editorial work was therefore constructed as a benchmark and measured on a compact desktop workstation, together with a deterministic baseline in which the same items are checked by a short program with no model. The panel is a schematic and shows no data.

# Materials and methods

## Tasks, guideline document, manuscripts, and ground truth

Eight tasks were measured (Figure 2). Guideline compliance (T1) asks a model to report every point at which a submitted manuscript departs from the journal's rules. Peer-review support (T2) asks for the points a reviewer would raise on a full paper. Reference formatting and Japanese-language rule queries (T3) test mechanical conversion and retrieval from the rules. Resistance to hallucination (T4) poses questions whose answers are absent from either the guidelines or the manuscript, so that the correct response is to state that the information is not available. Figure and table checking (T5) supplies images. Scientific self-consistency (T6) requires reconciling numbers stated in different sections. Category-conditional rules (T7) submits a manuscript under a category for which some rules are exempt. Reference verification with tools (T8) requires live calls to Crossref and the NLM Catalog.

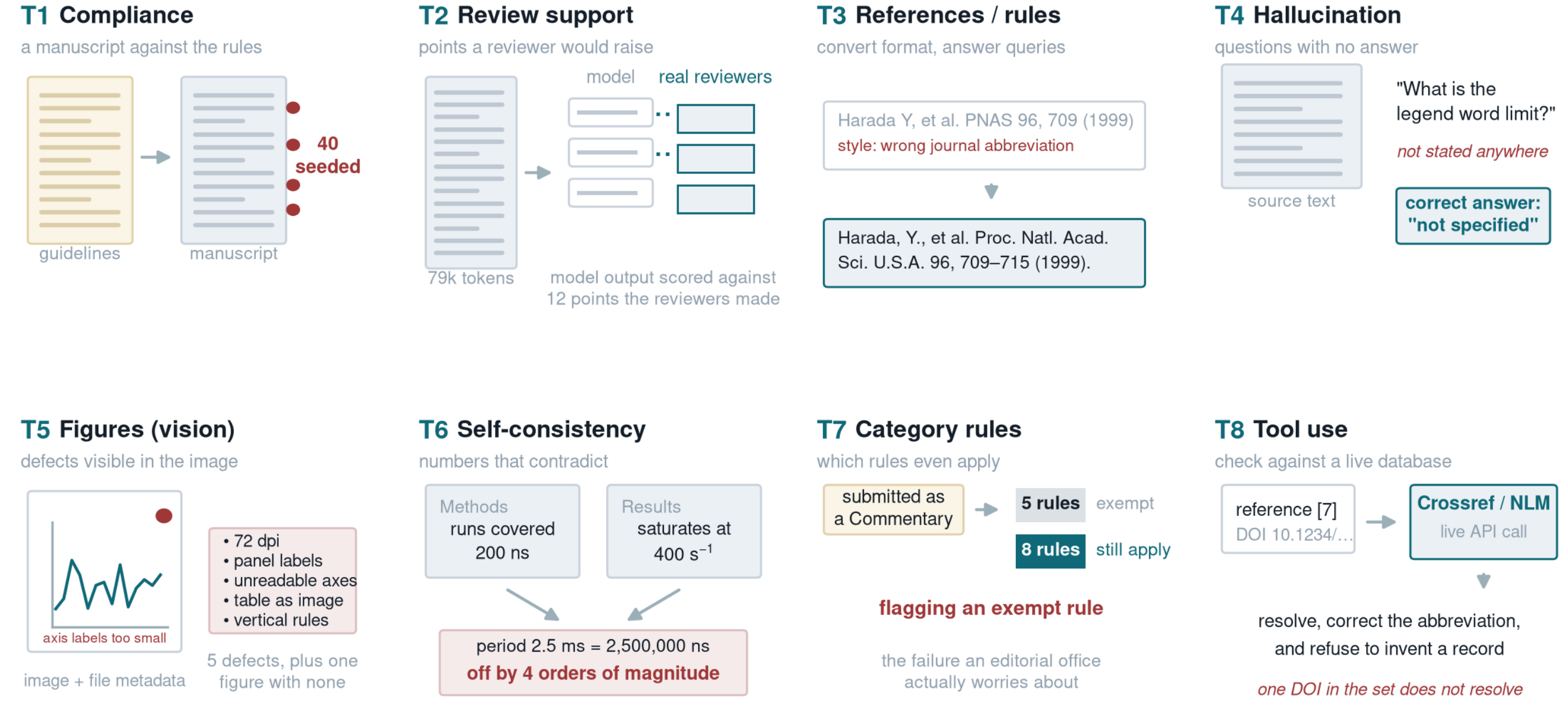

**Figure 2** The eight editorial tasks. T1 checks a manuscript against the journal's own Instructions for Authors, with 40 seeded violations. T2 generates review points and scores them against 12 points drawn from three published reviews. T3 converts reference formatting and answers rule queries. T4 poses questions whose answers are absent from the source, so that the correct response is a refusal. T5 supplies a figure image together with the file metadata a submission system reports. T6 requires reconciling numbers stated in different sections. T7 submits a manuscript under a category for which some rules are exempt, where the failure mode is flagging an exempt rule. T8 requires live calls to Crossref and the NLM Catalog, with one reference whose DOI does not resolve. The panel is a schematic and shows no data.

The guideline document underlying T1, T3, T4, and T7 is the Instructions for Authors of Biophysics and Physicobiology (https://www.biophys.jp/biophysics_and_physicobiology.html), version marked "Revised 21 August, 2026" and retrieved on that date (5,092 words), redistributed as `data/guidelines/` in the accompanying repository (see Data availability) with the permission of the Editor-in-Chief. A real rule set was used to preserve features of journal guidelines that would be difficult to reproduce synthetically, including numeric limits, formatting prescriptions, category-conditional exemptions, and descriptive prose that does not specify a rule. The document was converted to plain text by `harness/make_clean_guidelines.py` (see Data availability). The script exits with a non-zero status if any required section, including the version line, is missing from the converted text.

Three primary manuscripts were composed for this study in the style of a Regular Article in this journal, with invented authors, affiliations, and content: MS-A (2,832 words) for T1, MS-C for T6, and MS-E for T7. Two additional controls were derived from these manuscripts by script, giving five manuscripts in total. The manuscripts are available in `data/manuscripts/`, and the seeded defects and ground truth in `data/groundtruth/`, both in the accompanying repository (see Data availability). MS-A contains 40 guideline violations, together with 6 distractors — compliant text that resembles a violation — and 3 items excluded from scoring as artifacts of constructing a self-consistent manuscript. MS-C contains 10 internal scientific inconsistencies. MS-E is submitted under a category for which 5 rules are exempt and 8 apply. The two controls were derived rather than written independently so that the only differences were the seeded ones: MS-B is MS-A with every violation corrected, and MS-D is MS-C made consistent. On both controls, the correct output is an empty list.

The 40 violations in MS-A are classified as 25 EASY, decidable at a single location such as an abstract over its word limit, and 15 HARD, requiring two locations to be reconciled, an ordering judgment, or noticing an absence such as a table that is never cited. The classification is recorded item by item, with the reason for each, in `data/groundtruth/MS-A_difficulty.json` in the accompanying repository (see Data availability), so that the split can be recomputed directly. The classification was intended to distinguish local rule application from tasks requiring cross-referencing, counting, ordering, or detection of an absence.

The remaining tasks use item sets rather than whole manuscripts; all are available in `data/` in the accompanying repository (see Data availability). T2 used a real preprint and its three published public reviews, from which 12 distinct substantive points were extracted. One point was marked as a consensus point, and Reviewer 3 described it as "perhaps our biggest critique". T3 comprises 8 references, written by us in styles in which submissions might plausibly arrive, to be converted into the form prescribed by the Instructions for Authors, together with 10 questions in Japanese about those instructions. These questions test retrieval from a long English document and the ability to answer in Japanese. T4 comprises 8 questions about the guidelines, of which 4 have an answer and 4 do not, together with 5 questions about elements that do not exist in MS-A. In both sets, the correct response to an unanswerable item is to state that the information is unavailable; inventing a plausible answer is the failure the task is designed to detect. T8 comprises 8 references to be checked against Crossref and the NLM Catalog, including one with a DOI that does not resolve and two that disagree with the records to which they resolve. The T5 stimuli are four matplotlib-generated figures, three of which contain five defects between them while the fourth is compliant. The T5 prompt supplies both the image and the file attributes that a submission system would report. Consequently, only three of the five defects strictly require vision, and the T5 score is an upper bound on visual capability.

Ground truth was verified before use by `harness/verify_t6.py` and `harness/verify_t7.py` in the accompanying repository (see Data availability). These scripts perform 31 arithmetic checks on MS-C and 21 on MS-E, 52 in total, each recomputing a seeded defect from the manuscript's own stated numbers. The verification identified a genuine violation in MS-B, the nominally compliant control: Figure 4 was cited before Figure 3. The control was corrected and the affected conditions were re-run.

## Models and the tasks each could run

Twenty open-weight models from 3.3 to 81 GB were measured (Table 1). Which tasks a model ran was governed by the capabilities reported in the runtime's model metadata, but not only by them. Seventeen models ran the text tasks; all 12 models carrying a vision encoder ran T5; and 15 of the 18 models supporting tool calls ran T8. The three tool- capable models absent from T8 — both `qwen3-vl` variants, which were included for the figure task — were not run on it, so their absence is an omission rather than a capability limit.

**Table 1** Models evaluated. Size is the on-disk weight size reported by the runtime. MoE denotes a mixture-of-experts architecture, in which only a subset of the parameters is used for any one token; "Active" gives that subset per token, which governs how much must be read from memory during generation. V and T mark a vision encoder and tool-calling support.

| Model | Size | Type | Active | V | T |
|---|---|---|---|---|---|
| gemma3:4b | 3.3 GB | dense | 4B | ✓ | |

| Model | Size | Type | Active | V | T |
|---|---|---|---|---|---|
| gemma4 | 9.6 GB | dense | | ✓ | ✓ |
| gpt-oss:20b | 13 GB | MoE | | | ✓ |
| mistral-small | 14 GB | dense | | | ✓ |
| magistral | 14 GB | dense | | | ✓ |
| qwen3.6:27b | 17 GB | dense | 27B | ✓ | ✓ |
| qwen3.8:27b | 17 GB | MoE | | ✓ | ✓ |
| qwen3-vl:30b-a3b-instruct | 19 GB | MoE | 3B | ✓ | ✓ |
| qwen3.6:35b-a3b-q4_K_M | 23 GB | MoE | 3B | ✓ | ✓ |
| GLM-4.7-Flash Q8_0 | 31 GB | MoE | | | |
| qwen3.6:35b-a3b-q8_0 | 38 GB | MoE | 3B | ✓ | ✓ |
| nemotron | 42 GB | dense | 42B | | ✓ |
| command-r-plus | 59 GB | dense | | | ✓ |
| qwen3-vl:30b-a3b-instruct-bf16 | 62 GB | MoE | 3B | ✓ | ✓ |
| gpt-oss:120b | 65 GB | MoE | | | ✓ |
| glm-4.5-air:q4 | 67 GB | MoE | 12B | | |
| llama4:scout | 67 GB | MoE | | ✓ | ✓ |
| qwen3.6:35b-a3b-bf16 | 71 GB | MoE | 3B | ✓ | ✓ |
| qwen3.5:35b-a3b-bf16 | 71 GB | MoE | 3B | ✓ | ✓ |
| qwen3.5:122b-a10b-q4_K_M | 81 GB | MoE | 10B | ✓ | ✓ |

Two models support neither vision nor tool calling and so appear in neither T5 nor T8. `gemma3:4b` and the two `qwen3-vl` variants took part in T5 only, leaving 17 for the text tasks.

## Prompts, inference parameters, and scoring

Every condition was run at temperature 0 and at random seed 42. Fifteen conditions over seven models were additionally repeated at seeds 43 and 44 to check reproducibility (Table 2); the models leading each task were chosen for repetition, so that the ordering at the top rests on more than one measurement. Scores were not averaged over seeds, and each seed is reported as its own condition. For each task, `num_ctx` was set to the smallest context window that accommodated both the full prompt and the maximum permitted output: 131,072 for T2, 65,536 for T1, T6, and T7, 32,768 for T3, T4, and T8, and 16,384 for T5. T2 required the largest window because its prompt spans 70,286 to 78,983 tokens depending on the tokenizer. An earlier setting of 65,536 left no room for output and caused the runtime to evict the beginning of the paper while the model was generating its answer, invalidating nine conditions. `num_predict` was 24,576 for T1 to T4, T6, and T7, 16,384 for T8, and 8,192 for T5. Every request had the same wall-clock limit of 3,600 s. Runs exceeding this limit were recorded as timeouts.

Model-default `top_p`, `top_k`, and `min_p` were left unchanged and are inert under greedy decoding. Greedy decoding was nearly, but not exactly, reproducible: of the fifteen repeated conditions, twelve scored identically at all three seeds and three differed by one violation (Table 2).

The prompts are `harness/prompts.py` in the accompanying repository (see Data availability), one builder per task. All of them instruct the model to report only what the guidelines require, to answer "not specified" where the guidelines are silent, and to return its findings as JSON.

For T1, three prompt structures were compared while holding the guidelines and the manuscript constant. **Free-form** supplies both documents and asks for every violation in a single call. **Checklist** supplies the same documents with the guidelines decomposed into eight areas taken from the document's own section headings, again in a single call; the decomposition supplies structure, not answers. **Area-split** issues one call per area and concatenates the results. The remaining tasks use a single structure each, matched to what the task requires: T2 supplies the full text of the preprint and asks for the points a reviewer would raise; T3, T4, and T7 supply the guidelines together with the items to be judged; T5 supplies each figure as an image with its file attributes; T6 supplies the manuscript alone, since consistency is internal to it; and T8 supplies the references together with two callable functions and runs the model's tool calls against Crossref and the NLM Catalog, returning the results for the model to use.

Model outputs were matched against task-specific ground truth using scoring rules that differed by task. Models return JSON findings with `location`, `problem`, and `rule`; only the first two participate in matching, because `rule` quotes the journal's wording and produces spurious matches. On T1, T6, and T7 each ground-truth item carries anchor phrases specific enough that a finding about a neighbouring issue cannot be credited to it, and matching decides independently for each item whether any finding reports it. An item counts as detected when a finding contains one of its anchors and either that anchor is at least twelve characters long or the finding also contains one of the item's supporting keywords. Judging items independently rather than assigning findings one-to-one avoids crediting the wrong item when two share vocabulary and avoids penalising models that bundle related defects into one finding. T2 is matched one-to-one instead, by descending keyword overlap, because its twelve reference points are distinct criticisms and one statement by a model should not be credited to several of them. T3, T4, and T8 are not matched by keyword at all: each expected field is checked for exact presence, so a converted reference is correct only if the journal abbreviation, volume, pages, year, and DOI form all appear as specified. On T5 a defect counts as detected when the model's description of it contains one of that defect's keywords, which is the most permissive rule in the suite and makes T5 scores an upper bound in this respect as well as in its use of file metadata. Where a model contributes several conditions to one task — through a different prompt structure, a different reasoning effort, or a repeated seed — the figures relating score to model size and rank use that model's highest score on the task rather than its mean. This choice estimates each model's best observed capability rather than its average performance across the configurations tested.

T3 and T4 each yield two scores rather than one, and are reported as two measures throughout rather than pooled. Their item sets differ in material and in the ability they load on: T3's 8 references test mechanical conversion into a prescribed format, whereas its 10 Japanese questions test retrieval from a long English document and generation in another language; T4's 8 questions are drawn from the Instructions for Authors, whereas its 5 concern elements absent from MS-A. Models dissociate across each pair, and in opposite directions, so a pooled score would be set by the larger item set. On T3, `nemotron` converts none of the 8 references but answers 8 of the 10 Japanese questions, while `gpt-oss:20b` converts 4 references and answers none of the questions; pooled into an 18-item score these become 8 and 4 of 18, an ordering fixed by the 10-item set alone and silent about the conversion, on which the ranking is the other way round. On T4 the two sets agree for 14 of the 17 models, but glm-4.5-air:q4 answers all 8 guideline questions correctly while denying none of the 5 absent elements, and nemotron does the reverse.

Scores are related to model size by Spearman rank correlation rather than by Pearson, because the sample is small and a single model can dominate a product-moment coefficient: on the figure task the Pearson value over 11 models is +0.41, an apparent size effect deriving entirely from one 3.3 GB model of an older generation scoring 1/5, which turns negative (−0.31) when that model is excluded.

Keyword matching was used only as a prefilter. Every unmatched finding was written to results/t1_adjudicate.jsonl in the accompanying repository (see Data availability) and adjudicated by Claude Code, whose decisions H.O. reviewed. An unmatched finding may be a real violation that was not seeded, which is how the defect in MS-B surfaced. Reported false-positive counts are post-adjudication.

**Table 2** Violations detected at each random seed, for the fifteen conditions measured at more than one. MS-A carries the 40 seeded violations; MS-B is the compliant control, where the number is false positives. Twelve of the fifteen are identical across all three seeds and the other three differ by one, so the orderings reported here do not turn on the choice of seed.

| Model | Size | Prompt | Manuscript | seed 42 | seed 43 | seed 44 |
|---|---|---|---|---|---|---|
| qwen3.5:122b-a10b-q4_K_M | 81 GB | checklist | MS-A | 36 | 36 | 36 |
| qwen3.5:35b-a3b-bf16 | 71 GB | free-form | MS-A | 27 | 27 | 27 |
| qwen3.6:35b-a3b-bf16 | 71 GB | checklist | MS-A | 30 | 30 | 30 |
| qwen3.6:35b-a3b-bf16 | 71 GB | area-split | MS-A | 27 | 27 | 27 |
| qwen3.6:35b-a3b-bf16 | 71 GB | area-split | MS-B | 2 | 2 | 2 |
| gpt-oss:120b | 65 GB | free-form, low | MS-A | 10 | 10 | 11 |
| gpt-oss:120b | 65 GB | free-form, low | MS-B | 0 | 0 | 0 |
| qwen3.6:35b-a3b-q8_0 | 38 GB | free-form | MS-A | 26 | 26 | 26 |
| qwen3.6:35b-a3b-q8_0 | 38 GB | free-form | MS-B | 0 | 0 | 0 |
| qwen3.6:35b-a3b-q8_0 | 38 GB | area-split | MS-A | 32 | 32 | 32 |
| qwen3.6:35b-a3b-q8_0 | 38 GB | area-split | MS-B | 1 | 1 | 1 |

| Model | Size | Prompt | Manuscript | seed 42 | seed 43 | seed 44 |
|---|---|---|---|---|---|---|
| `qwen3.6:27b` | 17 GB | checklist | MS-A | 30 | 30 | 30 |
| `qwen3.8:27b` | 17 GB | free-form | MS-A | 29 | 29 | 30 |
| `qwen3.8:27b` | 17 GB | checklist | MS-A | 33 | 33 | 33 |
| `qwen3.8:27b` | 17 GB | checklist | MS-B | 1 | 1 | 0 |

### Deterministic baseline

To establish how much of the compliance task requires a language model at all, the same 40 violations were evaluated with a program that uses no model. The program was measured on the same manuscripts and under the same conditions as the language models. `harness/deterministic_check.py` in the accompanying repository (see Data availability) uses regular expressions and arithmetic only, and reads the same plain text the models were given. It performs no parsing of Word or LaTeX structure, which a production pipeline would have, so its score is a lower bound on what a scripted layer can reach. The baseline allows the contribution of the models to be stated as a difference rather than assumed.

### Workstation and inference stack

Measurements were made on an NVIDIA DGX Spark, a compact desktop workstation: a GB10 Grace Blackwell SoC (compute capability sm_121a), a 20-core Arm CPU (10 Cortex-X925 and 10 Cortex-A725), and 119 GiB of unified CPU/GPU memory with approximately 273 GB/s of bandwidth, running Ubuntu 24.04.3 (kernel 6.14.0-1013-nvidia) with CUDA 13.0 and driver 580.95.05. Because memory is unified there is no discrete VRAM, and the conventional GPU instrumentation reports `Memory-Usage: Not Supported`; all memory figures here come from `free` and per-process resident set size. Unified memory also removes the constraint that weights fit a discrete graphics card, allowing models from 3.3 to 81 GB to be evaluated on a single machine. Two-node measurements used a second identical machine connected by ConnectX-7 running remote direct memory access (RDMA) over Converged Ethernet (RoCEv2), and llama.cpp's RPC backend. RDMA lets one machine read the other's memory without involving its processor; activation was confirmed in the server log rather than assumed.

Inference used ollama. Two versions were in use: 0.17.6 as a system service, and 0.32.15 installed in user space because upgrading the system service required privileges that were unavailable. `qwen3.8:27b` and `gemma4` were released during the study and require the newer runtime, so they ran on 0.32.15 while the remaining models ran on 0.17.6. This is an uncontrolled factor. The harness, the graders, and the analysis are Python 3.11.14; they use only the standard library except for the figures, which use matplotlib and NumPy.

### Measurement of inference speed and memory usage

Inference proceeds in two phases: *prefill*, in which the model reads the prompt, processing all of its tokens together, and *decode*, in which it writes the answer one token at a time. Prefill throughput measured during task execution is contaminated by the runtime's prompt cache, which produced an uninformative apparent throughput of 518,067 tokens/s because a prompt the runtime has already seen is not read again. Speed was therefore measured in separate runs, each preceded by unloading the model so that no cache survives, and on passages cut from published papers rather than on the task prompts, so that no passage is presented twice. Peak memory is the maximum observed during execution, sampled at five context lengths from 4k to 128k to separate weight footprint from KV-cache growth. The two-node comparison is a throughput microbenchmark and not one of the eight tasks: 44 measurements of prefill and decode rate for two models, at prompt lengths of 4,096, 12,288, 32,768, 71,230, and 131,072 tokens, on one node and on two nodes over each of RDMA and TCP.

Every condition run under the configuration above is reported, including those that returned no usable output and those that exceeded the wall-clock budget of 3,600 s per request: a model that cannot complete the task within a budget applied equally to all of them is a result of the measurement rather than a condition to be set aside.

### AI Scientist system that carried out the work

The benchmark construction, implementation, execution, analysis, and drafting were performed by Claude Code version 2.1.238 (Anthropic) running the model `claude-opus-5`, directed by the author. "We" in this paper denotes the author team named above, which includes that system; where the text describes a judgment in the first person — which defects to seed, how to classify them, which of a model's unmatched findings are real — the judgment was made by Claude Code and reviewed by H.O., in the manner set out under Author contributions. This is recorded here as part of the method rather than only in the acknowledgements, because it bears on how the work should be read: several intermediate conclusions reported in the Discussion were produced by Claude Code and later corrected, in some cases by Claude Code itself on re-examination and in others after the author challenged them. Everything the system produced was reviewed by the author, who corrected what the review found, directed how the argument is presented, and takes responsibility for the content; what that review covered is set out under Author contributions. The measurements themselves are independent of that system: they are produced by the scripts in the repository and can be re-run without any assistance from a language model.

---

## Results

### Detection of seeded guideline violations by the baseline and the models (T1)

Practical adoption depends on how much of the editorial work a locally hosted model can perform and how large that model must be. To separate tasks that require a model from those that do not, the 40 seeded violations in MS-A were evaluated with the deterministic baseline, with each model under its best prompt, and with the baseline and model in combination. The deterministic baseline detected 31 of the 40 violations, the strongest single model detected 36, and the combination detected all 40 (Table 3). The baseline raised no false positive on the compliant control and took under a tenth of a second and no memory; the model raised one and took 722 s and 81 GB.

**Table 3** Contribution of each layer to the 40 seeded guideline violations (T1, MS-A). The deterministic checker is 220 lines of regular expressions and arithmetic and uses no model. The external lookup resolves a journal name to its Index Medicus abbreviation through Crossref and the NLM Catalog; it is listed for completeness, since within this item set the best model already reports that violation, and the lookup is what supplies the correct form (Results, tool use). "Best model" is `qwen3.5:122b-a10b-q4_K_M` under the checklist prompt. Its one false positive on the compliant control is a distractor: text written to resemble a violation without being one. The final row is the union of detections from the two

layers measured separately, not the result of running an integrated pipeline. Time and memory are for a single manuscript.

| Layer | Detected | False positives | Time | Memory |
|---|---|---|---|---|
| Deterministic checker (regex + arithmetic) | **31/40** | **0** | **< 0.1 s** | **0** |
| External lookup (Crossref, NLM Catalog) | 1 item, already covered | 0 | one API call | 0 |
| Best model alone (`qwen3.5:122b-a10b-q4_K_M`, checklist) | 36/40 | 1 | 722 s | 81 GB |
| **Union of the two** | **40/40** | **1** | 722 s | 81 GB |

The checker and the model failed on largely different violations. The checker detected four violations that the best model missed, each settled by a regular expression or by arithmetic: tables not numbered in order of citation, a graphical-abstract caption of 134 words against a limit of 100, a citation written as "(1, 2)" where the journal prescribes brackets, and a database accession given without the prescribed format or its DOI. `qwen3.5:122b-a10b-q4_K_M` under the checklist prompt reaches the nine violations the checker cannot decide, which are those requiring a judgment about wording or about what a statement covers. The union of their detections covers all 40 violations, which is an arithmetic combination of two separately measured layers rather than an evaluated pipeline. The complementarity is not, however, a property of models in general, since no violation was detected by every model condition: each of the 40 was missed by some condition, which argues against a single-model design independently of which model is chosen. The two detect largely the same violations — 27 of the 40 are found by both — so what is complementary on this item set is not their detections but their remaining errors. Relative to the deterministic baseline, the model contributes nine additional detections rather than forty, at approximately 7,000 times the wall-clock time and with an 81 GB memory footprint.

One violation shows that the two are not interchangeable. MS-A cites only its own Table 2 in its text, so its tables are not numbered in order of appearance; the checker settles this by comparing the citation order against the numbering, and all 62 model conditions missed it (Figure 3). For this item, the deterministic layer was not merely a cheaper route to a result that a model could also reach; it was the only successful route we observed. The figure also shows a bimodal pattern in task completion: 12 of the 62 conditions detected no violation at all. Three exhausted the 24,576-token generation budget on reasoning and returned no answer, two exceeded the time budget, five returned output from which no findings could be recovered, and two emitted barely a hundred tokens.

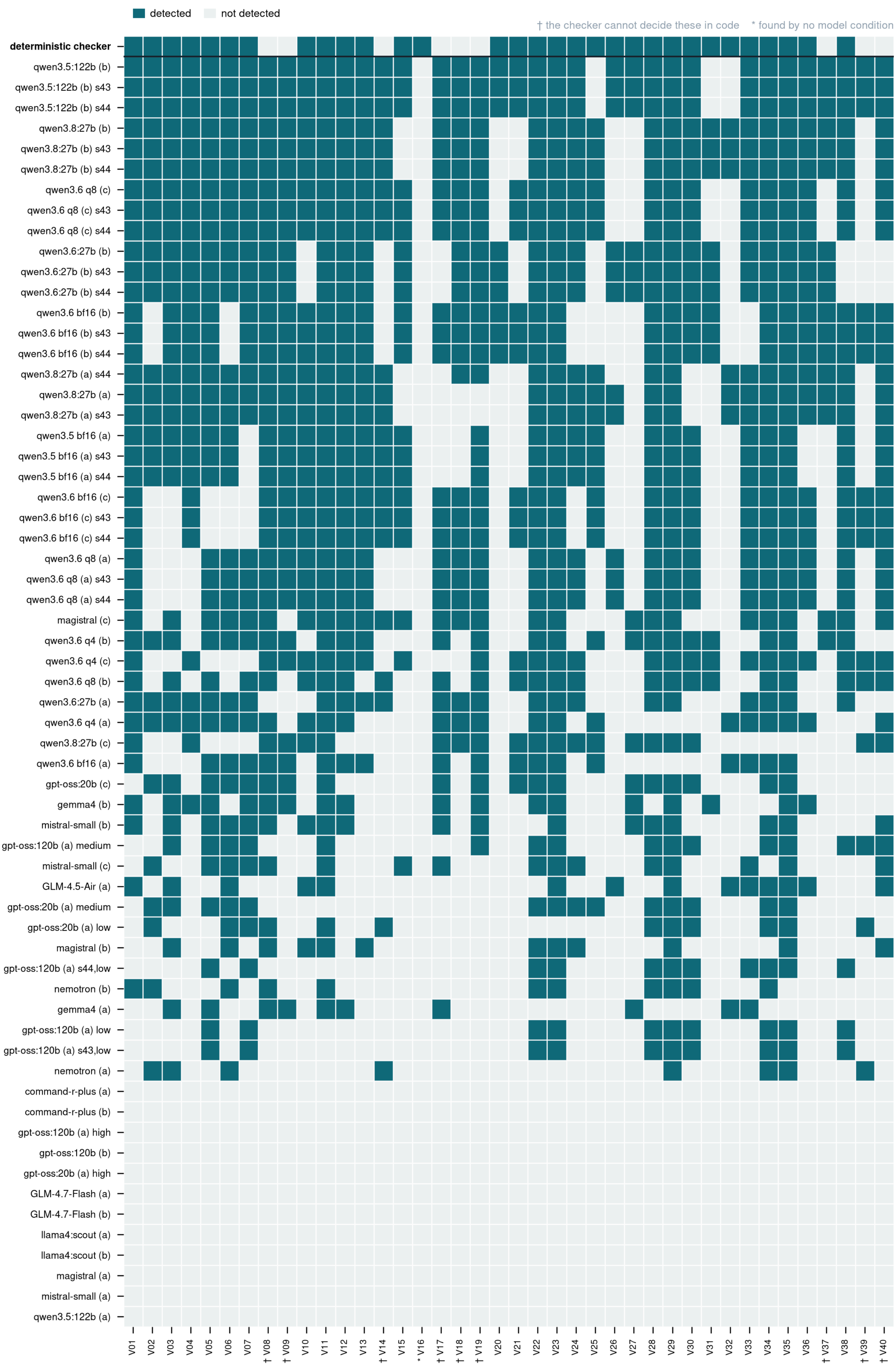

detected
not detected
† the checker cannot decide these in code
* found by no model condition
deterministic checker
qwen3.5:122b (b)
qwen3.5:122b (b) s43
qwen3.5:122b (b) s44
qwen3.8:27b (b)
qwen3.8:27b (b) s43
qwen3.8:27b (b) s44
qwen3.6 q8 (c)
qwen3.6 q8 (c) s43
qwen3.6 q8 (c) s44
qwen3.6:27b (b)
qwen3.6:27b (b) s43
qwen3.6:27b (b) s44
qwen3.6 bf16 (b)
qwen3.6 bf16 (b) s43
qwen3.6 bf16 (b) s44
qwen3.8:27b (a) s44
qwen3.8:27b (a)
qwen3.8:27b (a) s43
qwen3.5 bf16 (a)
qwen3.5 bf16 (a) s43
qwen3.5 bf16 (a) s44
qwen3.6 bf16 (c)
qwen3.6 bf16 (c) s43
qwen3.6 bf16 (c) s44
qwen3.6 q8 (a)
qwen3.6 q8 (a) s43
qwen3.6 q8 (a) s44
magistral (c)
qwen3.6 q4 (b)
qwen3.6 q4 (c)
qwen3.6 q8 (b)
qwen3.6:27b (a)
qwen3.6 q4 (a)
qwen3.8:27b (c)
qwen3.6 bf16 (a)
gpt-oss:20b (c)
gemma4 (b)
mistral-small (b)
gpt-oss:120b (a) medium
mistral-small (c)
GLM-4.5-Air (a)
gpt-oss:20b (a) medium
gpt-oss:20b (a) low
magistral (b)
gpt-oss:120b (a) s44,low
nemotron (b)
gemma4 (a)
gpt-oss:120b (a) low
gpt-oss:120b (a) s43,low
nemotron (a)
command-r-plus (a)
command-r-plus (b)
gpt-oss:120b (a) high
gpt-oss:120b (b)
gpt-oss:20b (a) high
GLM-4.7-Flash (a)
GLM-4.7-Flash (b)
llama4:scout (a)
llama4:scout (b)
magistral (a)
mistral-small (a)
qwen3.5:122b (a)
V01
V02
V03
V04
V05
V06
V07
† V08
† V09
V10
V11
V12
V13
† V14
V15
* V16
† V17
† V18
† V19
V20
V21
V22
V23
V24
V25
V26
V27
V28
V29
V30
V31
V32
V33
V34
V35
V36
† V37
V38
† V39
† V40
seeded violation

**Figure 3** Per-violation detection on MS-A. Columns are the 40 seeded violations; rows are the detectors — the deterministic checker above the rule, then every model condition, ordered by how many violations it detected. Filled cells are detections, as in the key. A dagger on a column label marks a violation the checker cannot decide in code; an asterisk marks the one violation that no model condition detected. Conditions are labelled by model and prompt structure (a, free-form; b, checklist; c, area-split), with the random seed or the reasoning effort added where one model contributes several conditions.

Partitioning the nine violations the checker cannot decide indicates that fewer than half of them require a model at all: four require reading a statement and judging what it covers, three are layout properties a document parser could recover, one is a lookup an authority file settles (Results, tool use), and one is a section-placement rule. The classification is recorded item by item, with the reason for each, in `data/groundtruth/MS-A_unreachable_partition.json`. On this manuscript, therefore, four of the forty violations appeared to require semantic interpretation beyond the deterministic mechanisms considered here. A language model is one way to supply that interpretation; whether it is the only one these measurements do not establish, since a document model, a symbolic pipeline or a human might also reach them. The first question can accordingly be answered as follows: a 17 GB model reaches 33 of the 40 where the largest model in the set reaches 36, and four of the forty lie beyond the deterministic layer for reasons of interpretation rather than of format. What the measurement does not settle is how the split would fall on manuscripts we did not write, where the mix of violations may differ.

## Effect of model weight size on score (T1–T8)

Hardware selection often assumes that capability increases with model size. If that assumption holds, model choice largely reduces to a resource constraint; if it does not, weight size is not a reliable procurement criterion. Each model is represented here by its highest score on the measure, so the comparison is between the best configurations tested for each; because the number of configurations tried differs between models, that statistic favours the models tried in more of them, and the analysis is weaker than one holding the configuration fixed. The relation was therefore examined across all eight tasks rather than on the compliance task alone, by correlating each model's weight size with its normalised score on every measure it ran. T3 and T4 each contribute two measures, labelled a and b, because their item sets test different abilities on different material and models dissociate across them in opposite directions (Materials and methods); the eight tasks therefore give ten panels (Figure 4).

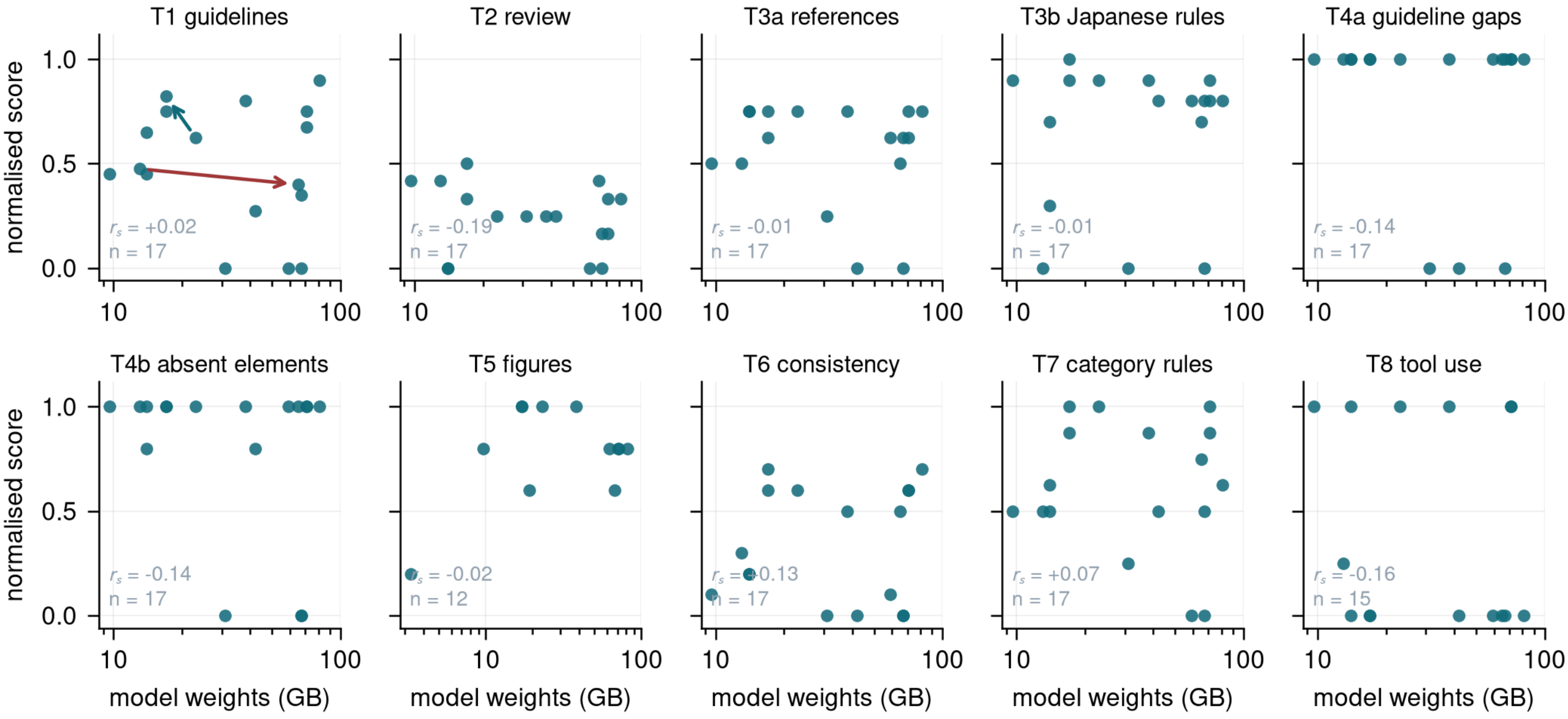

**Figure 4** Model weight size against the highest normalised score the model reached on the measure shown, over every model that ran it. Because the number of configurations tried differs between models, this statistic favours the models tried in more of them. There are ten panels for the eight tasks: T3 and T4 each contribute two. Their item sets test different abilities on different material — reference conversion against Japanese-language retrieval for T3, questions drawn from the Instructions for Authors against questions about elements absent from MS-A for T4 — and models dissociate across each pair in opposite directions, so a pooled score would be set by the larger set. On T3, `nemotron` converts none of the 8 references but answers 8 of the 10 Japanese questions, while `gpt-oss:20b` converts 4 and answers none. The Spearman rank coefficient and the number of models plotted are given in each panel: 17 models on T1, T2, T3a, T3b, T4a, T4b, T6, and T7, 12 on T5, which only vision-capable models could run, and 15 on T8, which only tool-capable models could run. No significance test is reported for the coefficients, and the absence of a monotonic relation is an absence of a detectable association rather than evidence of none. Two arrows in the T1 panel mark comparisons in which the larger or older model scores lower. The red arrow runs from `gpt-oss:20b` at 13 GB and 19 of 40 to `gpt-oss:120b` at 65 GB and 16 of 40, a five-fold increase in weight within one family for three fewer violations. The teal arrow runs from `qwen3.6:35b-a3b-q4_K_M` at 23 GB and 25 of 40 to `qwen3.8:27b` at 17 GB and 33 of 40, a later generation that is smaller and scores higher. No other panel carries arrows; these two comparisons are singled out because they hold family or lineage fixed, which the correlation across all models does not.

Using each model's best observed configuration, no task showed a consistent monotonic ordering by weight size. Spearman coefficients lie between −0.19 and +0.13, and the sign is as often negative as positive. The observed ordering does not follow size even where the largest model leads: on T1 the 81 GB model reaches 36 of 40, but the next three at or above 60 GB reach 30, 27, and 16, below the 33 of a 17 GB model, and within the `gpt-oss` family the 65 GB member scores 16/40 against 19/40 for its 13 GB sibling.

Figure checking (T5) showed the same lack of a size-dependent ordering. Among the 12 vision-capable models the four perfect scores are at 17, 17, 23, and 38 GB, while the five models at 62–81 GB score 3 or 4 and the largest false-positive count on the compliant figure came from a 67 GB model. Comparing the 17–38 GB band with the ≥62 GB band gives a difference of 0.80 points at an exact permutation *p* of 0.119.

For the second question, across the best configurations tested we observed no consistent monotonic association between weight size and score on any task measured here. This result supports no detectable size advantage rather than a size disadvantage: the data are consistent with no effect, and the sample is too small to establish the reverse. The models tested also span one moment in a moving field, so the finding is about this population rather than about scaling in general.

## Model rank within each task (T1–T8)

A deployment must still choose a model on some basis, so it matters whether an ordering is a property of the model or of the pairing between model and task. Ranks here are taken from the same best-configuration scores, and the prompt variants were not run on every model — the area-split form on 7 of 17 — so part of any reordering may follow from which configurations a model was given. Each model was therefore ranked within each measure and the ranks compared across them (Figure 5).

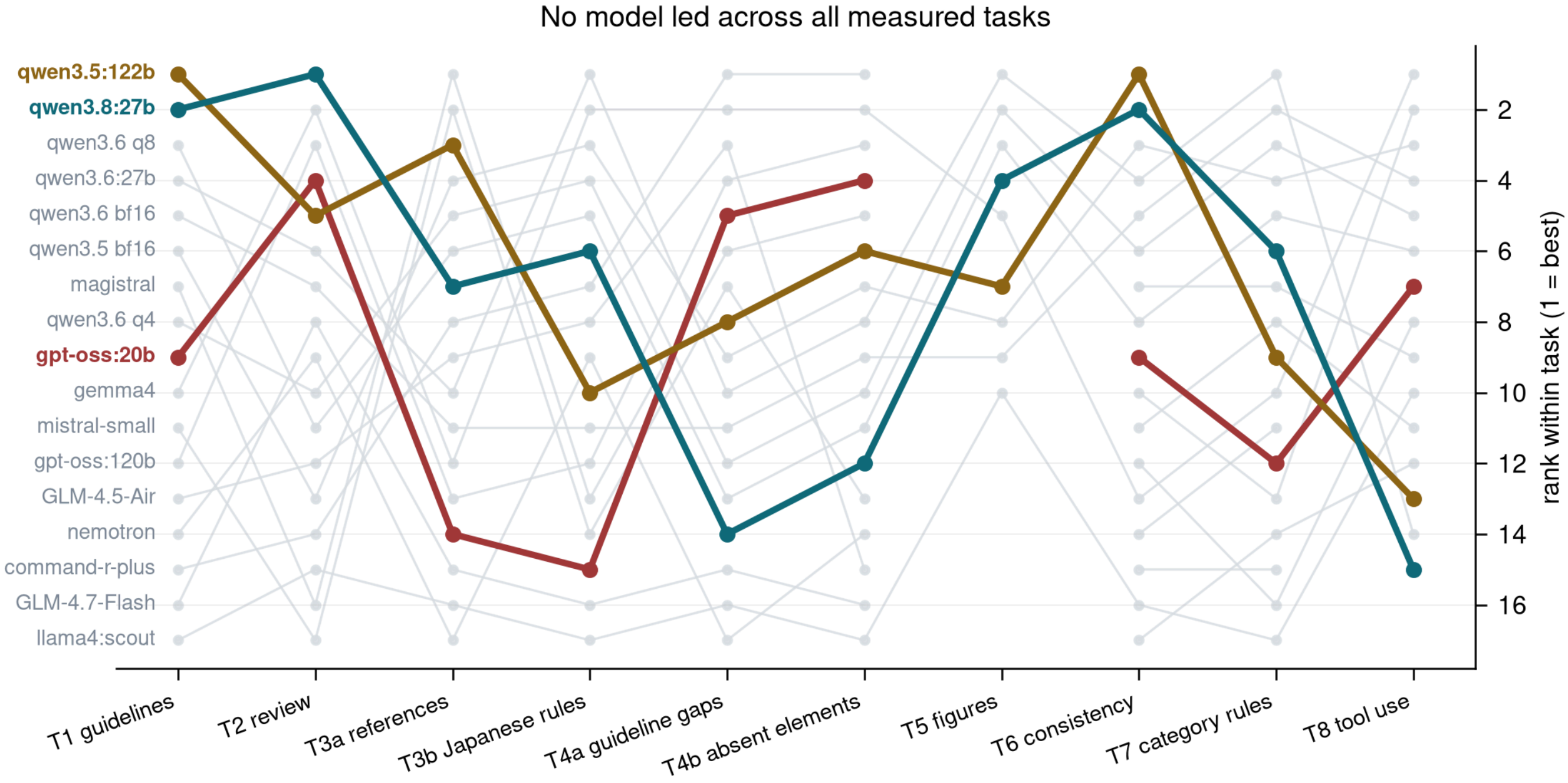

**Figure 5** Rank within each measure, model by model, over the same ten measures as Figure 4. Every line is named at its left-hand end, in the colour of the line. All 17 lines begin at T1, where the ranks run from 1 to 17 and are therefore distinct, so the names read down the left margin in T1 order. Three models are drawn in colour and the remaining 14 in grey. Lines break at measures a model did not run. The three vision-only models that ran T5 alone are omitted, since a single point is not a rank trajectory.

No model led across all measured tasks. The model that leads guideline compliance (T1), `qwen3.5:122b-a10b-q4_K_M`, is 5th of 17 on peer-review points (T2) and returns none of T8's four journal abbreviations correctly; `qwen3.8:27b` is 2nd on T1 and 1st on T2 and returns none of them either; and the models that return all four include `gemma4`, which ranks 10th of 17 on T1. Assigning models to roles by task rather than choosing one selects an 81 GB model for T1, a 17 GB model for T2, a 17 GB model for figure checking (T5), and a 9.6 GB model for reference lookups (T8): four roles that no single model fills. Held resident together at a 128k context those four require 136 GiB, more than this machine has; loaded one at a time the largest of them needs 84.8 GiB, which fits. These measurements establish that a model can lead one task and fail another outright, but do not determine why performance differs so sharply between them. These results indicate substantial variation in observed model rank across tasks, which suggests that a deployment may benefit from task-specific model selection; how far that carries beyond this benchmark is limited by the single preprint behind T2 and the five defects behind T5.

### Effects of prompt structure, quantization, and model generation on score (T1)

Because model size did not account for the observed differences, we next examined factors that could change performance without requiring larger hardware. Two were task-side changes: how the task was posed and whether the model was given a tool for operations that are lookups rather than inferences. We also examined two model-side changes within one family: higher-precision quantization and a later model generation. Prompt structure moved most models measured under both forms on T1, but not all of them and not always upward: of the 14 models measured under both, 9 detected more under the checklist, 2 detected fewer, and 3 detected nothing under either. The largest gains belong to models that return nothing at all under free-form prompting and therefore start from zero; among the 7 that produce output under both forms the shift ranges from -2 to +11 violations. Where the gain falls is not uniform either (Table 4).

**Table 4** Detection by difficulty class under the two prompt structures, for two models that respond to the checklist differently (T1, MS-A). For `qwen3.8:27b` the gain is on the violations decidable at a single location and the harder class does not improve; for `qwen3.6:27b` the gain is on the harder class. The direction of the effect is therefore a property of the model as well as of the prompt.

| Model | Prompt | EASY (25) | HARD (15) | Total | Wall time |
|---|---|---|---|---|---|
| `qwen3.8:27b` | Free-form | 19 | 10 | 29 | 674 s |
| | **Checklist** | **24** | 9 | **33** | 605 s |
| `qwen3.6:27b` | Free-form | 18 | 5 | 23 | 1,140 s |
| | **Checklist** | **21** | **9** | **30** | 1,305 s |

Figure 6 places these comparisons on one axis. They are not interventions of one kind: the deterministic baseline is a different computational layer rather than a change of model, and the others are single comparisons within one model family or between two generations, so the panel reports observed differences and not interchangeable effects. The newer-generation model in the matched comparison detected eight additional violations while occupying 11 GiB less, and implementing the deterministic baseline recovered 31 at no memory cost at all. Decomposing the prompt recovered four for the model shown. Within the `qwen3.6:35b` family neither quantization step helped: raising q4 to q8 lost one violation for 13.8 GiB more memory, and q8 to bf16 lost seven for a further 30.4 GiB. Whether that holds for other families was not tested.

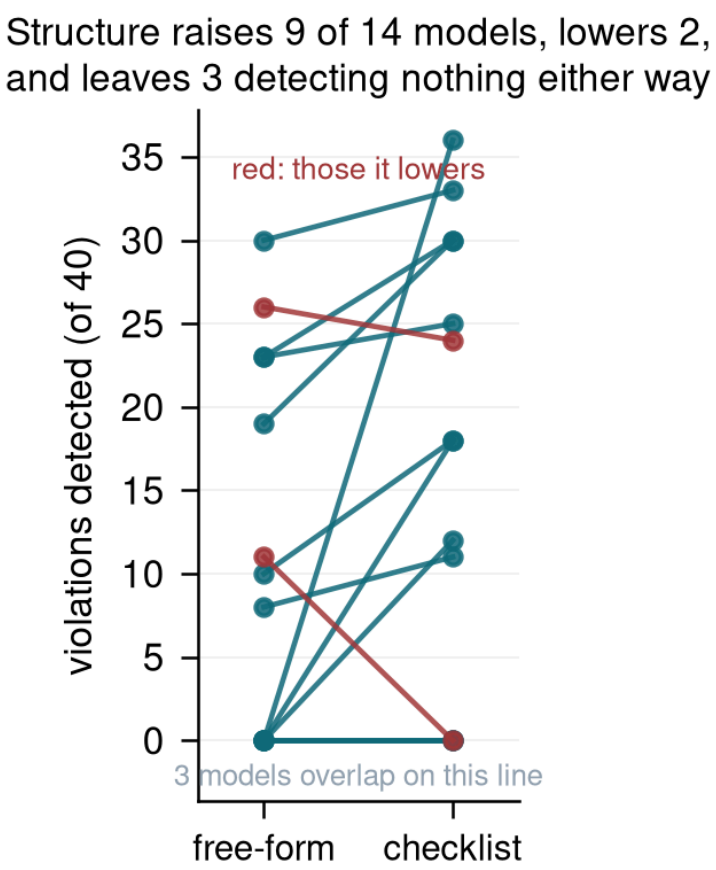

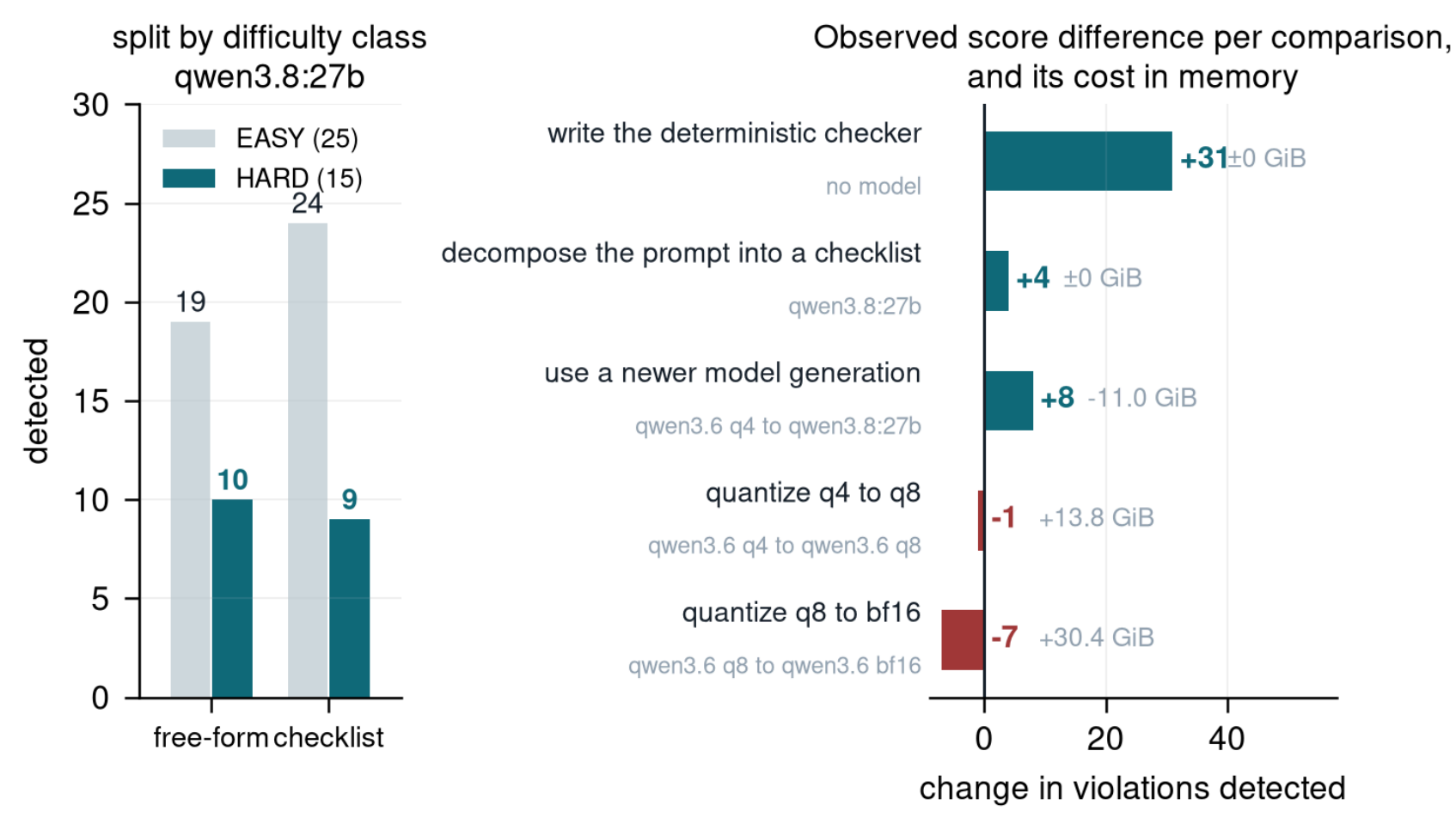

**Figure 6** What moved detection on T1. Left: guideline violations detected under free-form prompting and under the checklist, one line per model at matched reasoning effort; red marks the two models the checklist lowers, and the three models that detect nothing under either form lie on one coincident line at zero, marked as such. Centre: the same shift resolved by difficulty class for `qwen3.8:27b`. Right: the observed score difference for each comparison, annotated with its cost in memory; the second line of each row names the models compared, from `data/score_change_comparisons.json`. The checklist row is `qwen3.8:27b` under the two prompt forms, the generation row is `qwen3.6:35b-a3b-q4_K_M` against `qwen3.8:27b`, and the two quantization rows are steps within `qwen3.6:35b-a3b`; the deterministic checker uses no model. The comparisons are not interventions of one kind — the deterministic checker is a different computational layer, and the others are single comparisons within one model family or between two generations — so the panel reports differences rather than interchangeable effects.

Tool access showed a similar distinction between possessing a capability and using it effectively. Rendering `PNAS` as `Proc. Natl. Acad. Sci. U.S.A.` is a lookup rather than an inference, and in T1, where no tool is available, only 17 of the 62 conditions report it. In T8 all 15 tool-capable models are given Crossref and the NLM Catalog, and 6 return the correct abbreviation for all four of the references that carry one, the smallest at 9.6 GB; of the 9 that do not, three never issue a tool call and six issue between 2 and 19 calls and still answer wrongly, the largest model in the set among them. Access to an authority file therefore resolves the lookup only when a model both invokes the tool and uses the returned information correctly.

Prompt structure also affected whether some models completed T1 successfully. One 14 GB model replicated a single finding 358 times in free-form, running 37 minutes to reach 4/40, but completed in 156 seconds at 19/40 under the checklist and reached 22/40 area-split without looping; another completed only under structure, and a third collapsed under it, so a model cannot be assessed from one prompt format. These measurements do not establish why prompt structure helps some models and hurts others. On the third question, prompt structure produced score differences of up to 11 violations within matched models and required no additional hardware, whereas no consistent score ordering by weight size was observed and neither quantization step improved the score.

## Peak memory and the prefill–decode split (T1 and T2)

We next examined the machine itself. Hardware requirements depend on two quantities: the memory occupied by a model at paper-length context and the inference phase that dominates wall time. Both were measured directly.

Memory capacity was not the binding constraint for a 17 GB model, which holds a 128k-token context for +1.1 GiB over its 4k footprint, reaching 17.2 GiB. How much the KV cache adds is architecture-dependent by more than an order of magnitude: a 67 GB model requires +48.8 GiB over the same range, reaching 113.8 GiB, and the 81 GB model that leads on T1 requires only +4.4 GiB, reaching 84.8 GiB. Three models exceed 96 GiB at a 128k context, and they are the three that the task measurements also place at the bottom: one degenerates into repetition, one fails to complete the compliance task within the time budget, and one detects no violation at all. On this set, the models that most strain the memory of the machine are the models least usable for the work.

The dominant inference phase depended on the task (Figure 7). The compliance run shown there, `qwen3.8:27b` on MS-A under the checklist, reads 13,029 tokens and writes 19,219, because the model reasons at length before enumerating findings with quoted evidence: prefill takes 16.9 s and decode 580.6 s, so decode is 97% of the wall time. Every model condition that produced more than 5,000 output tokens — 55 of the 88 T1 conditions — spent 94–100% of wall time in decode. Prefill dominated only in conditions that produced almost no output, such as one model that emitted 95 tokens. Peer-review support (T2) shifts a much larger share of the time into prefill, but not for every model: the same model reads 78,983 tokens and writes 3,505, and decode still accounts for 59% of the wall time, while for the two `gpt-oss` conditions in Figure 7 prefill takes the majority, at 74% and 69%. The phase split is therefore consistent with a potentially important role for memory bandwidth rather than capacity on T1, but it does not identify the binding hardware constraint. Bandwidth was not varied, so this is a reading of the phase split rather than a manipulation, and whether the same relation holds on hardware with several times the bandwidth cannot be established from a single platform.

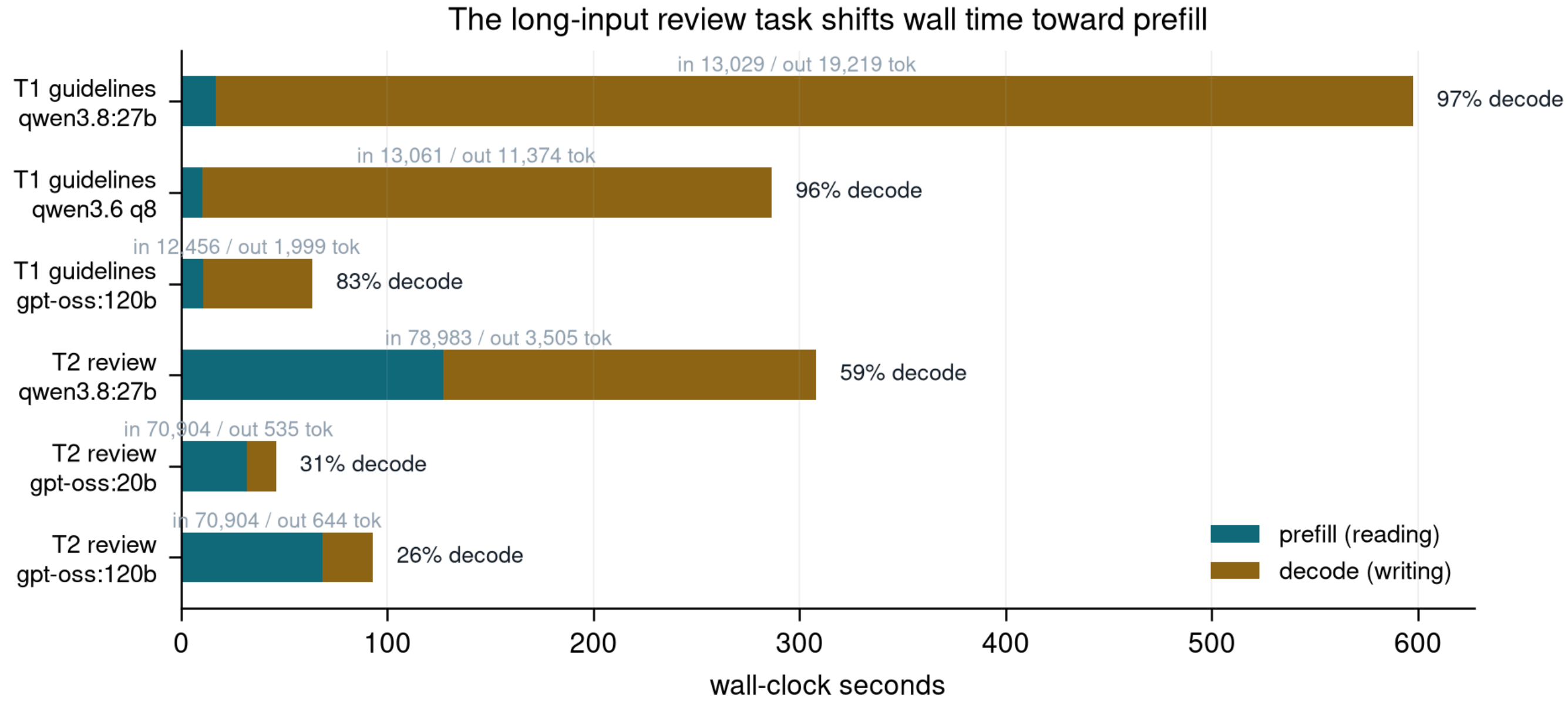

**Figure 7** Wall-clock time split into prefill and decode for the two use cases — guideline compliance (T1) and peer-review support (T2) — with input and output token counts.

## Prefill and decode throughput on one and two nodes

Finally, we tested whether adding a second node improved inference throughput. This was measured as a throughput microbenchmark rather than by repeating the eight tasks, because splitting a model across nodes changes token-processing rates rather than task content. The resulting throughput measurements are related back to T1 at the end of this subsection. Splitting a model across two nodes over RDMA raises prefill throughput by 1.5–1.8× over a 32-fold range of prompt length and a fourfold range of model size (Figure 8): 1.64–1.77× for a 63.06 GiB model at five prompt lengths from 4,096 to 131,072, and 1.54–1.65× for a 15.65 GiB model at three. The gain reflects compute parallelism rather than memory relief: the 15.65 GiB model, under no memory pressure on a 119 GiB node, still gains 1.5–1.6×. Decode moved far less than prefill, and by an amount that depended on the model: 1.5–1.9% for the 15.65 GiB model, but 3.5–17.5% for the 63.06 GiB one, where two-node decode was flat at 24.8–25.0 tokens/s against 21.3–24.0 tokens/s on one node. Decode is sequential, so the second node cannot parallelise it; splitting the weights does halve what each node reads per token, which would raise decode on the model large enough for that to matter, but the reads were not measured directly. Over TCP instead of RDMA, the same split runs at 0.46–0.54× the throughput of a single node, making two nodes slower than one. The RDMA-to-TCP throughput ratio is 3.3–3.8× at the two prompt lengths measured over TCP, showing that interconnect performance is a defining part of the multi-node configuration. The prefill gain has little effect on end-to-end T1 runtime: the T1 compliance run above takes 605 s of which 16.9 s is prefill, so a 1.5–1.8× prefill speedup saves 6–8 s, or about 1%. Scaling this platform out therefore accelerates a phase that contributes little to T1 wall time. On this platform and inference stack, a second node is most likely to benefit workloads dominated by long-input prefill.

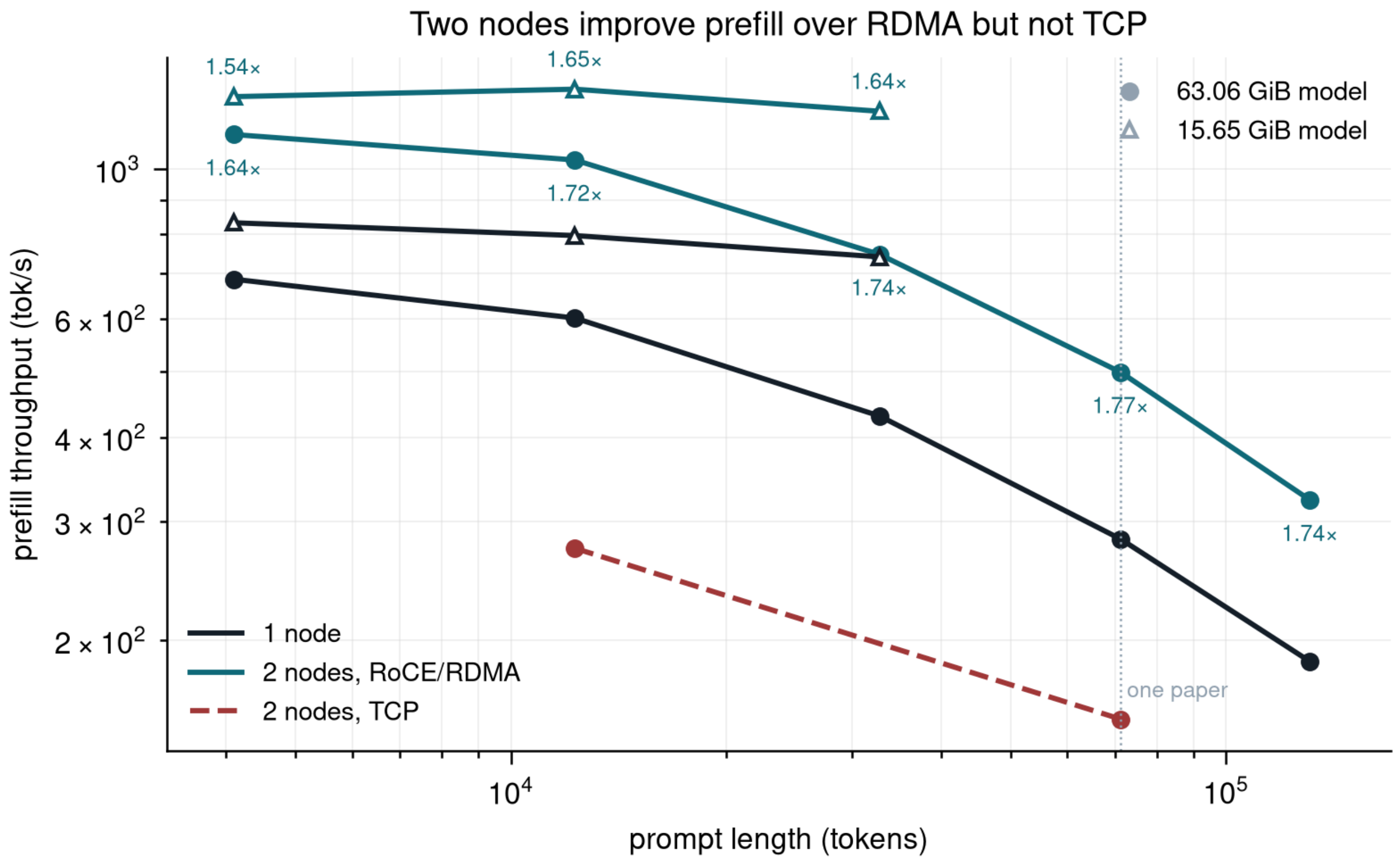

**Figure 8** Prefill throughput against prompt length for a single node, two nodes over RDMA, and two nodes over TCP. Both measured models are plotted: colour carries the configuration and marker shape the model, filled circles for GLM-4.5-Air at Q4_K_XL and 63.06 GiB, measured at five prompt lengths and on both interconnects, and open triangles for a 15.65 GiB model measured at three prompt lengths over RDMA only. Ratios beside the RDMA points are relative to one node at the same prompt length and the same model. Only these two interconnects were tested, so the comparison bears on RDMA against TCP on this stack rather than on distributed inference in general.

---

## Discussion

We set out to measure how much of a journal's editorial work a locally hosted model can perform, whether that capability tracks model size, and what other factors affect the result. The strongest model detected 36 of the 40 seeded guideline violations and a 17 GB model detected 33; the best model on peer-review support recovered 6 of the 12 points raised by human reviewers. Across the best configurations tested, however, we observed no consistent monotonic association between weight size and score on any of the eight tasks. What did change the result was the layer the work was assigned to, and after that the generation of the model and the way the task was posed; within the one model family in which quantization was varied, raising it did not help.

Three quarters of the seeded guideline violations were recovered by a program using no model, with no model-memory requirement and in under a tenth of a second. The violations they resolved are the cross-referencing, counting, and formatting ones, and they include four that the highest-scoring model missed. The result with the broadest practical implication is therefore not a particular model score, but an ordering of computational layers.

This ordering is consistent with established practice in editorial screening. Statistical consistency checking with statcheck is already used during peer review at several journals [30], and automated assessment of reporting-guideline compliance has been integrated into submission platforms for editorial use. Such systems combine conditional random fields with pattern matching to assess criteria including randomisation, blinding, power analysis, and resource identifiers [31,32]. Our measurements suggest a similar architecture in which the language model occupies the final and narrowest computational role: deterministic checks first, external tools for tasks that are lookups rather than inferences, a language model for the remaining semantic judgments, and a human for what none of these layers can reliably resolve. Applying a language model before deterministic checks have been implemented spends inference on work that does not require it. For an editorial office, implementing the deterministic layer primarily costs programmer time rather than inference resources, while also establishing the baseline against which subsequent model performance can be judged. Without that baseline, the contribution of the models in our experiment would have appeared to be 36 recovered violations rather than nine additional ones.

Some failures were largely unaffected by model capacity. An order-of-magnitude inconsistency spanning a unit conversion — a 200 ns simulation reported as showing saturation of a process whose period is 2.5 ms, a discrepancy of nearly five orders of magnitude — was found by 1 of the 17 models, a 38 GB one, and missed by the other 16 including every model larger than it. Another seeded inconsistency, a claim of isotropy contradicted by the manuscript's own figures, was found by none. The Index Medicus abbreviation of a journal name was recovered unaided by some models and not others, with no relation to size: 9 of the 17 models converted every journal name correctly on T3, and the failures included the largest model in the set. That task is a lookup rather than an inference, and attaching a bibliographic tool resolved it for the models that both called the tool and used what it returned. Both failures therefore point to the same conclusion as the layer ordering above: some editorial tasks are better assigned to mechanisms other than a language model.

Peer-review support produced a more qualified result. The best model recovered 6 of the 12 reference points in 315 s. The model recovered points that also appeared in the human reviews, but this experiment does not establish whether presenting such output to a reviewer improves the review: points the model raised that matched no reference point were logged for inspection rather than scored, so precision against unsupported criticism was not measured. For the point identified by one reviewer as the most important, 9 of 17 models detected it and 8 did not; four of those eight produced no usable output at all. These findings qualify the premise of the collection in which this paper appears, namely that AI scientists can contribute to research alongside human ones [33,34]. The contribution measured here is demonstrable, but narrower than that framing might suggest, and it is greatest when the model is combined with methods that are not themselves models. The surrounding infrastructure required for such workflows is already feasible: the Journal of Open Source Software has operated a bot-mediated editorial process since its first year [22].

The measurements also separate three hardware quantities that are easily conflated when a system is described only by its memory capacity. In our setting, capacity determined which models could be loaded. A single-role deployment of the 17 GB model occupies 17.2 GiB at a 128k context, well within a compact desktop workstation, and the strongest model on the compliance task occupies 84.8 GiB, which still fits. What does not fit is the best model for every task held resident at once, which would require 136 GiB on a machine with 119. Among the compliance-task conditions that produced more than 5,000 output tokens, 55 of 88, decode accounted for 94-100% of wall time, and the long-input review task moved a much larger share of that time into prefill, though decode still took the majority for two of the four conditions measured. Adding a second node raised prefill throughput by 1.5-1.8x, and raised decode by 1.5-1.9% for a 15.65 GiB model but by 3.5-17.5% for a 63.06 GiB one. Prefill is therefore the more compute-sensitive phase, and the size-dependence of the decode gain is consistent with decode being limited by the rate at which each node reads weights, which splitting the model across two nodes halves. Neither quantity was varied independently, so this is a reading of the phase behaviour rather than a measurement of either.

These quantities do not scale together. Characterising a machine by memory capacity alone therefore misses the phase behaviour that governs inference time on this workload. That behaviour is consistent with memory bandwidth being an important constraint, but bandwidth was not varied independently, so these measurements cannot identify it as the causal or dominant one. This also limits the generality of our speed measurements. The platform used here combines relatively strong compute with mid-range memory bandwidth. In the configurations measured, dense models decoded substantially more slowly than mixture-of-experts models with few active parameters: at a 4k prompt a 42 GB dense model reached 4.6 tokens/s against 43.3 for a same-generation mixture-of-experts model with 3B active parameters [35], consistent with decode being limited by the rate at which weights can be read from memory [36]. That gap narrows with prompt length, to 1.3-fold at 65k tokens. If decode is bandwidth-limited as prior work suggests, hardware with substantially higher memory bandwidth should reduce it. Higher-bandwidth hardware was not tested. The nearest measurement available here is the two-node split, which roughly doubles the aggregate rate at which weights can be read: it raised decode by 3.5-17.5% on the 63.06 GiB model and by 1.5-1.9% on the 15.65 GiB one, in the direction predicted and on the model large enough for the rate to bind, but without separating bandwidth from the compute the second node also adds. The ranking of architectures by speed could therefore change even if their ranking by task capability did not.

Several limitations constrain the interpretation of these results. First, the manuscripts were our own. Real submissions were unavailable, so the benchmark trades representativeness for exact ground truth. Absolute recall should therefore not be interpreted as expected field performance, although the within-benchmark comparisons between models and prompting conditions remain informative. Extending the benchmark to real submissions under confidentiality-preserving arrangements would test how closely these results transfer to editorial practice.

Second, the peer-review task is based on a single preprint from one subfield, and its 12 reference points were extracted by us from review prose. Whether the observed advantage of the best-performing model generalises remains unknown. A larger reference set spanning multiple journals and disciplines would provide a stronger test.

Third, most conditions were evaluated with a single random seed. Fifteen conditions over seven models were repeated at three seeds: twelve scored identically and three differed by one violation (Table 2). That is reassuring for the orderings reported here, but fifteen conditions are insufficient to establish confidence intervals. We therefore report no significance tests other than the permutation test used for the figure task. The figure task itself also does not isolate visual reasoning, because only three of the five seeded defects strictly require information from the image. A stimulus set in which every defect is purely visual would measure that capability more directly.

Several implementation factors also remain uncontrolled. Two runtime versions were used, and the top-scoring model ran on the newer version. A single-runtime replication would remove this potential confound. All inference was performed within one runtime family, so the throughput measurements are specific to that software stack and should be replicated with alternative runtimes. Similarly, the two-node experiments were microbenchmarks of prefill and decode throughput rather than executions of the complete task suite across two nodes; the reported end-to-end estimate was calculated from the measured phase throughputs.

One prompt variant was evaluated on only 7 of the 17 models because each condition required eight inference calls. No principal conclusion depends on this variant alone, but completing the comparison would provide a stronger test of the effect of prompt structure. Finally, the deterministic layer used here operates on plain text and does not parse document structure. A production implementation with access to the submitted file format could therefore recover several violations that we assigned to the model. This would likely narrow the model's incremental contribution further and is perhaps the most tractable extension of the present work.

## Conclusion

Locally hosted language models performed a subset of the editorial benchmark tasks on a compact desktop workstation of the class tested here, and across the best configurations tested capability did not increase consistently with model size. In guideline-compliance screening a deterministic plain-text checker recovered 31 of 40 seeded violations and the highest-scoring model 36, with the union of their detections covering all 40. Of the nine the checker could not decide, four were classified as requiring semantic interpretation and the rest as document structure, a lookup, or a placement rule. Peer-review support was more demanding, and the best model recovered only 6 of the 12 points raised by human reviewers.

These results argue against choosing an editorial model on weight size alone. Deterministic checks and external tools should handle tasks they can solve reliably, with language models reserved for work that requires semantic interpretation. This division of labour allows useful editorial support while retaining the confidentiality benefits of local deployment.

---

## Conflict of interest

The author declares no conflict of interest.

## Author team composed of AI scientist(s) and human scientist(s)

Haruka Ozaki, RIKEN Center for Biosystems Dynamics Research / University of Tsukuba; Claude Code version 2.1.238 (Anthropic), model claude-opus-5.

## Author contributions

H.O. posed the question, supplied the institutional context and the hardware, established the requirement that the study measure editorial work rather than compare specifications, directed the scope at each stage, and reviewed and revised the manuscript.

Claude Code designed the eight-task benchmark including the EASY/HARD classification; wrote the five manuscripts and seeded the 58 ground-truth defects; implemented the execution harness, the grader, the ground-truth verifiers, the deterministic control arm, and the two-node measurement scripts; executed and monitored all 313 conditions and the 44 two-node measurements; diagnosed the failure modes reported here; performed the analysis; produced the figures; and drafted the manuscript.

Every part of this work was produced in the first instance by Claude Code and subsequently reviewed by H.O.: the text, the figures, the tables, every number reported, the code, and the ground truth. Numbers were checked against the stored measurements, statements about the method against the implementation, and citations against the works cited, and what the review found to be wrong was corrected. Beyond matters of correctness, H.O. determined how the argument is organised, the structure and wording of the text, and the design of the figures, and required passages to be rewritten where nothing was factually wrong but the presentation was unclear, the register unsuited to a research article, or the reasoning hard to follow; much of the present text and several of the figures took their form that way. Several conclusions in this paper are corrections of earlier conclusions drawn by Claude Code, identified either by Claude Code on re-examination or in response to challenges from H.O.; these are documented in the Discussion rather than silently amended. H.O. takes responsibility for the content of the manuscript.

---

## Data availability

The benchmark, execution harness, ground-truth data, all 313 measured conditions, the figure-generation code, and a chronological working log that records predictions which proved wrong are available at `https://github.com/bioinfo-tsukuba/local-llm-editorial-benchmark`. The journal's Instructions for Authors are redistributed there with the permission of the Editor-in-Chief, in the version marked "Revised 21 August, 2026" as requested.

All tabulated numbers are generated from the stored results by `harness/make_report.py` and `harness/make_paper_figures.py`; none are transcribed by hand. Ground truth passed 52 independent arithmetic checks before any model was scored.

---

## Acknowledgements

The author thanks Yuichi Togashi, Ryota Yamada, and Haruki Nakamura for helpful discussions that motivated this study.

This work was supported by JST CREST (JPMJCR2551 to H.O.), the RIKEN TRIP Advanced General Intelligence for Science Program (AGIS) (to H.O.), and the Japan Society for the Promotion of Science (JSPS) KAKENHI Grant Number, JP26HP2004.

Claude Code version 2.1.238 (Anthropic), running the model `claude-opus-5`, was used as described under Author contributions. The work was carried out in a single continuous session between 21 and 29 August 2026, in which the model issued 1,694 assistant turns; the session transcript is the record of what was directed and what was produced. Claude Code performed the implementation, measurement, and analysis, and drafted this manuscript; it also produced several of the erroneous intermediate conclusions documented in the Discussion.

ChatGPT (OpenAI, model ChatGPT 5.6 Sol) was used separately to search the literature when assembling the references; every work it surfaced was then resolved independently against Crossref or arXiv, and its open-access status checked, before being cited. It was also used to improve the readability of the text and to check its grammar; the wording it proposed was accepted only where H.O. judged it to preserve the meaning. No other AI system was used in the preparation of this work. The models listed in Table 1 are the *objects* of measurement and had no part in writing it.

---